\documentclass[12pt]{article}

\newenvironment{sciabstract}{%
\begin{quote} \bf}
{\end{quote}}

\newcounter{lastnote}

\def\be{\begin{equation}}
\def\ee{\end{equation}}

\usepackage{scicite}
\usepackage{times}
\usepackage{graphicx}
\usepackage{caption}

\title{The Equation of State of a Low-Temperature Fermi Gas with Tunable Interactions}

\author
{N. Navon$^{\dag\ast}$, S. Nascimb\`ene$^\dag$, F. Chevy, C. Salomon \\
\\
\normalsize{Laboratoire Kastler Brossel, CNRS, UPMC, \'Ecole
Normale Sup\'erieure,}\\
\normalsize{24 rue Lhomond, 75231 Paris, France}\\
\\
\normalsize{$^\dag$These authors contributed equally to this work.}\\ 
\normalsize{$^\ast$To whom correspondence should be addressed; E-mail: navon@ens.fr.}
}

\date{}

\begin{document} 

\baselineskip24pt

\maketitle 

\begin{sciabstract}

Interacting fermions are ubiquitous in nature and understanding their thermodynamics is an important problem. We measure the equation of state of a two-component ultracold Fermi gas for a wide range of interaction strengths at low temperature. A detailed comparison with theories including Monte-Carlo calculations and the Lee-Huang-Yang corrections for low-density bosonic and fermionic superfluids is presented. The low-temperature phase diagram of the spin imbalanced gas reveals Fermi liquid behavior of the partially polarized normal phase for all but the weakest interactions. Our results provide a benchmark for many-body theories and are relevant to other fermionic systems such as the crust of neutron stars. 

\end{sciabstract}\newpage

Recently, ultracold atomic Fermi gases have become a tool of choice to study strongly correlated quantum systems because of their high controllability, purity and tunability of interactions \cite{inguscio2006ultracold}. In the zero-range limit, interactions in a degenerate Fermi system with two spin-components are completely characterized by a single parameter $1/k_Fa$, where $a$ is the $s$-wave scattering length and $k_F=(6\pi^2n)^{1/3}$ is the Fermi momentum ($n$ is the density per spin state). In cold atom gases the value of $|a|$ can be tuned over several orders of magnitude using a Feshbach resonance, this offers an opportunity to entirely explore the so-called BCS-BEC crossover, \emph{i.e.} the smooth transition from Bardeen-Cooper-Schrieffer (BCS) superfluidity at small negative values of $a$ to molecular Bose-Einstein Condensation (BEC) at small positive values of $a$ \cite{inguscio2006ultracold,leggett1980modern}. Between these two well-understood limiting situations $a$ diverges, leading to strong quantum correlations. The description of this system is a challenge for many-body theories, as testified by the large amount of work in recent years \cite{inguscio2006ultracold}. The physics of the BEC-BCS crossover is relevant for very different systems, ranging from neutron stars to heavy nuclei and superconductors. 

In the grand-canonical ensemble and at zero temperature, dimensional analysis shows that the Equation of State (EoS) of a two-component Fermi gas, relating the pressure $P$ to the chemical potentials $\mu_1$ and $\mu_2$ of the spin components can be written as
\begin{equation}
P(\mu_1,\mu_2,a)=P_0(\mu_1)h\left(\delta_1\equiv\frac{\hbar}{\sqrt{2m\mu_1}a},\eta\equiv\frac{\mu_2}{\mu_1}\right),\label{pressure_h}
\end{equation}
where $P_0(\mu_1)=1/15\pi^2(2m/\hbar^2)^{3/2}\mu_1^{5/2}$ is the pressure of a single-component ideal Fermi gas, $m$ is the atom mass and $\hbar$ is the Planck constant divided by $2\pi$. The indices 1 and 2 refer to the majority and minority spin components, respectively. From the dimensionless function $h(\delta_1,\eta)$, it is possible to deduce all the thermodynamic properties of the gas, such as the compressibility, magnetization or the existence of phase transitions; the aim of this paper is to measure $h(\delta_1,\eta)$ for a range of interactions ($\delta_1$) and spin imbalances ($\eta$) and discuss its physical content. $\delta_1$ is the grand-canonical analog of the dimensionless interaction parameter $1/k_Fa$.

\emph{In-situ} absorption images of harmonically trapped gases are particularly suited to investigate their EoS as first demonstrated at MIT \cite{shin2008des} and ENS \cite{nascimbene2009eos}. In the particular case of the grand-canonical ensemble, a simple formula relates the local pressure $P$ at a distance $z$ from the center of the trap along the $z$ axis to the doubly-integrated density profiles $\overline{n}_1$ and $\overline{n}_2$ \cite{ho2009opdtq}:
\begin{equation}
\label{Pn}
P(\mu_{1}(z),\mu_{2}(z),a)=\frac{m\omega_r^2}{2\pi}
\left(\overline{n}_1(z)+\overline{n}_2(z)\right).
\end{equation}
Here we define the local chemical potentials $\mu_{i}(z)=\mu_i^0-\frac{1}{2}m\omega_z^2z^2$, where $\mu_i^0$ is the chemical potential of the component $i$ at the bottom of the trap, assuming local density approximation. $\omega_r$ and $\omega_z$ are the transverse and axial angular frequencies of a cylindrically symmetric trap respectively, and  $\overline{n}_i(z)=\int\;n_i(x,y,z)\textrm{d}x\textrm{d}y,$ is the atomic density $n_i$ of the component $i$, doubly integrated over the transverse $x$ and $y$ directions. In a single experimental run at a given magnetic field, two images are recorded, providing $\overline{n}_1(z)$ and $\overline{n}_2(z)$ (see Fig.S4 in \cite{sciencesom}); the $z$-dependence of the chemical potentials then enables the measurement of $P$ along a curve in the $(\delta_1,\eta)$ plane \cite{sciencesom}. This method was validated in \cite{nascimbene2009eos} for the particular case of the unitary limit $a=\infty$. Deducing the function $h$ from the doubly integrated profiles further requires a precise calibration of $\omega_z$ and the knowledge of the central chemical potentials $\mu_i^0$ \cite{sciencesom}. 

Our experimental setup is presented in \cite{nascimbene2009pol}. We prepared an imbalanced mixture of $^6$Li in the two lowest internal spin states, at the magnetic field of $834~G$ (where $a=\infty$), and trapped it in a hybrid magnetic-optical dipole trap. We then performed evaporative cooling by lowering the optical trap power, while the magnetic field was ramped to the final desired value for $a$. The cloud typically contained
$N=2$~to~$10\times10^4$ atoms in each spin state at a temperature of $0.03(3)$ $T_F$, justifying our $T=0$ assumption \cite{sciencesom}. The final trap frequencies
are $\omega_z/2\pi\sim30$~Hz, $\omega_r/2\pi\sim1$~kHz. Below a critical spin population imbalance, our atomic sample consists of a fully-paired superfluid occupying the center of the trap, surrounded by a normal mixed phase and an outer rim of an ideal gas of majority component atoms \cite{shin2006observation,nascimbene2009pol,nascimbene2009eos}. 

\begin{figure}[h!t!]
\centerline{\includegraphics[width=\columnwidth]{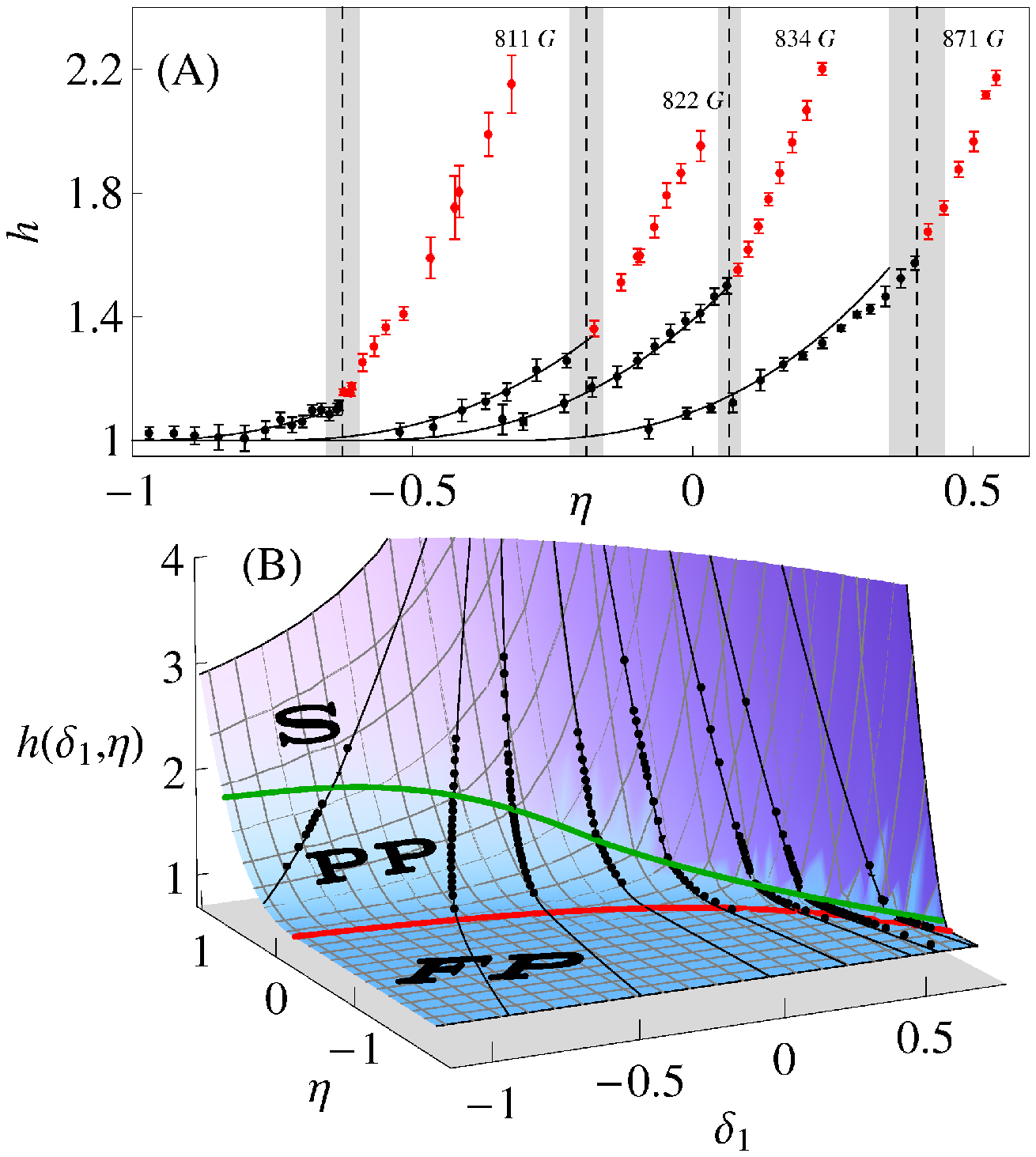}}
\caption{$h(\delta_1,\eta)$ of a zero-temperature two-component Fermi gas in the BEC-BCS crossover. (A): Samples of the data for different magnetic fields. The black (red) data points correspond to the normal (superfluid) phase, and are separated at $\eta_c(\delta_1)$ by a clear kink in the local slope of $h$. Solid black lines are the predictions of the polaron ideal gas model, Eq (\ref{ideal_gas}). The scattering length corresponding to each curve is (from left to right) : $(1.7, 3.4,\infty,-1.3)$ in units of $10^4$ $a_0$, where $a_0$ is the Bohr radius. (B): $h(\delta_1,\eta)$. The black dots are data recorded for each magnetic field value (as in Fig.\ref{fig_h}a). The black lines correspond to the parametric curves $(\delta_1(\eta),\eta)$ scanned by the density inhomogeneity in the harmonic trap \cite{sciencesom}. The red line is $A(\delta_1)$, the frontier between the fully polarized (FP) ideal gas $h=1$ and the normal partially polarized (PP) phase. The green line is $\eta_c(\delta_1)$ marking the phase transition between the normal and superfluid (S) phases. The surface is the parametrization of $h(\delta_1,\eta)$ given in the text.
}\label{fig_h}
\end{figure}

For a given magnetic field, 10 to 20 images are taken, leading after averaging to a low-noise EoS along one line in the $(\delta_1,\eta)$ plane. Measurements at different magnetic fields chosen between 766~G and 981~G give a sampling of the surface $h(\delta_1,\eta)$ in the range $-1<\delta_1<0.6$ and $-2<\eta<0.7$ (Fig.\ref{fig_h}). Let $A(\delta_1)$ be the limiting value of the ratio of chemical potentials $\mu_1(z)/\mu_2(z)$ below which the minority density vanishes. At fixed $\delta_1$ and $\eta<A(\delta_1)$, $h(\delta_1,\eta)$ represents the EoS of an ideal Fermi gas of majority atoms and is equal to $1$. For $\eta>A(\delta_1)$, it slowly rises and corresponds to the normal mixed phase, where both spin components are present. At a critical value $\eta=\eta_c(\delta_1)$, the slope of $h$ abruptly changes \cite{sciencesom}, the signature of a first-order phase transition from the normal phase (for $A<\eta<\eta_c$) to a superfluid phase with a lower chemical potential imbalance ($\eta>\eta_c$). We notice that the discontinuity is present for all values of $\delta_1$ we investigated, and this feature is more pronounced on the BEC side.

\begin{figure}[h!t!]
\centerline{\includegraphics[width=\columnwidth]{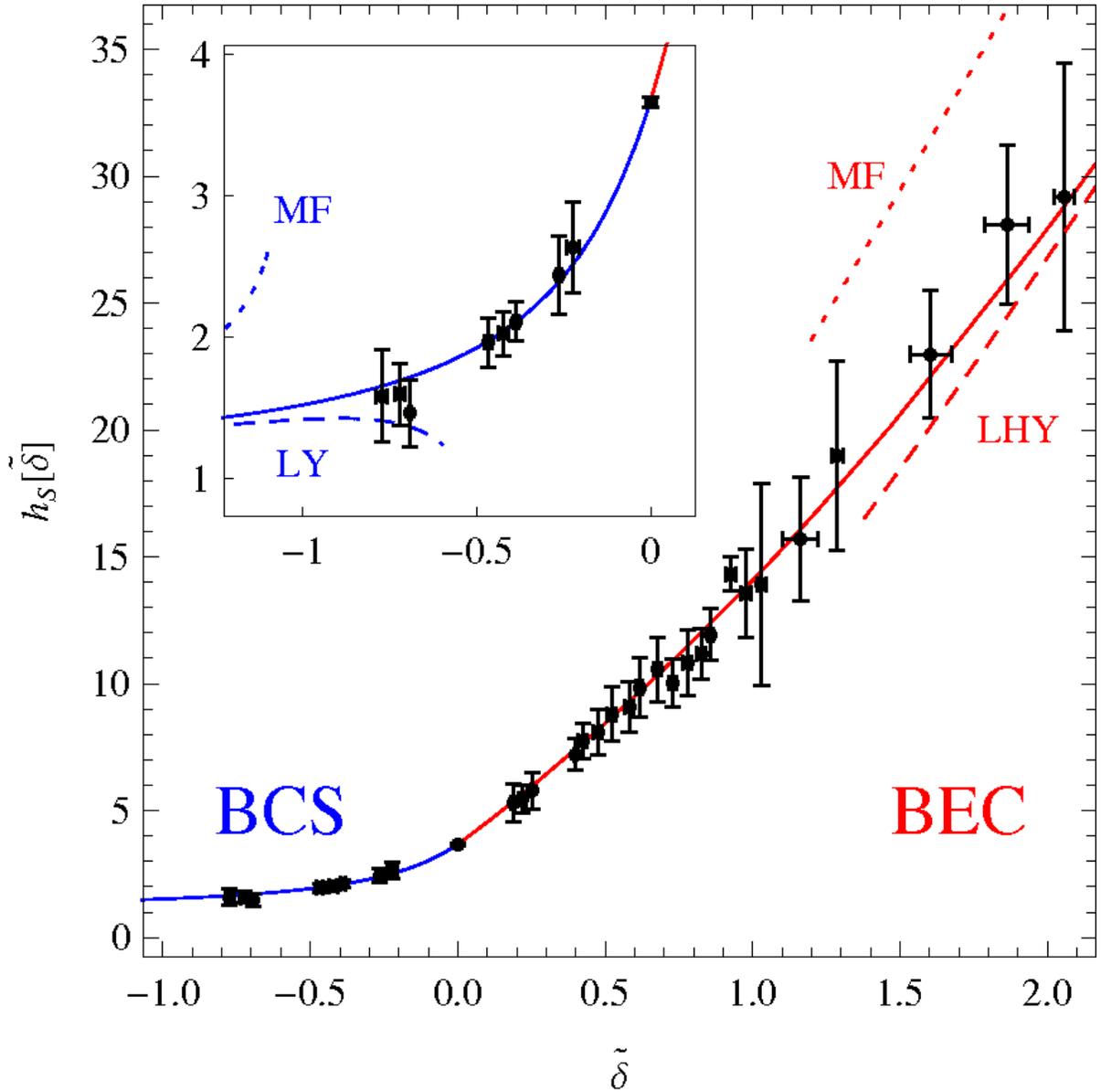}}
\caption{$h_S(\widetilde{\delta})$ of the $T=0$ balanced superfluid in the BEC-BCS crossover (black dots). The blue solid line is the fit $h_S^{\mathrm{BCS}}(\widetilde{\delta})$ on the BCS side of the resonance, the red solid line is the fit $h_S^{\mathrm{BEC}}(\widetilde{\delta})$ on the BEC side (see text). The dotted (dashed) red line is the mean-field (Lee-Huang-Yang) theory  \cite{footnoteBEC}. 
Inset: Zoom on the BCS side. The dotted (resp. dashed) blue line is the EoS including the mean-field (resp. Lee-Yang) term. The systematic uncertainties on the $x$ and $y$-axis are about $5$ $\%$. The errors bars represent the standard deviation of the statistical uncertainty.}\label{fig_h0}
\end{figure}

Let us first consider the EoS of the superfluid phase, $\eta>\eta_c$. Each of our \emph{in-situ} images has, along the $z$-axis, values of the chemical potential ratio $\eta(z)=\mu_{2}(z)/\mu_{1}(z)$ both lower and greater than $\eta_c$. In the region where $\eta(z)>\eta_c$ the doubly-integrated density difference $\bar{n}_1(z)-\bar{n}_2(z)$ is constant within our signal-to-noise ratio (see Fig.S4). This is the signature of equal densities of the two species in the superfluid core, \emph{i.e.} the superfluid is fully paired. Using Gibbs-Duhem relation $n_i=\frac{\partial P}{\partial\mu_i}$, equal densities $n_1=n_2$ imply that $P(\mu_1,\mu_2,a)$ is a function of $\mu$ and $a$ only, where $\mu\equiv(\mu_1+\mu_2)/2$. For the balanced superfluid, we then write the EoS symmetrically:
\begin{equation}
P(\mu_1,\mu_2,a)=2P_0(\widetilde{\mu})h_S\left(\widetilde{\delta}\equiv\frac{\hbar}{\sqrt{2m\widetilde{\mu}}a}\right).\label{eq_h0}
\end{equation}
In order to avoid using negative chemical potentials, we define here $\widetilde{\mu}=\mu-E_b/2$, where $E_b$ is the molecular binding energy $E_b=-\hbar^2/ma^2$ for $a>0$ (and $0$ for $a\leq0$). $h_S(\widetilde{\delta})$ is then a single-variable function. It fully describes the ground state macroscopic properties of the balanced superfluid in the BEC-BCS crossover and is displayed in Fig.\ref{fig_h0} as black dots. 

In order to extract relevant physical quantities, such as beyond mean-field corrections, it is convenient to parametrize our data with analytic functions. In this pursuit, we use Pad\'e-type approximants \cite{sciencesom}, interpolating between the EoS measured around unitarity and the well-known mean-field expansions on the BEC and BCS limits. The two analytic functions, $h_S^{\mathrm{BCS}}$ and $h_S^{\mathrm{BEC}}$ are respectively represented in blue and red solid lines in Fig.\ref{fig_h0} and represent our best estimate of the EoS in the whole BEC-BCS crossover. 

First, on the BCS side $\widetilde{\delta}<0$, $h_S^{\mathrm{BCS}}$ yields the following perturbative expansion of the energy in series of $k_Fa$:
\[
E=\frac{3}{5}NE_F\left(1+\frac{10}{9\pi}k_Fa+0.18(2)(k_Fa)^2+0.03(2)(k_Fa)^3+\ldots\right),
\]
where $N$ is the total number of atoms, $E_F$ is the Fermi energy and where by construction of $h_S^{\mathrm{BCS}}$, the mean-field term (proportional to $k_Fa$) is fixed to its exact value $10/9\pi$. We obtain beyond mean-field corrections up to $3^\mathrm{rd}$ order. The term proportional to $(k_Fa)^2$ agrees with the Lee-Yang \cite{leeyang1957many,diener2008quantum} theoretical calculation $4(11-2\log2)/21\pi^2\simeq0.186$. The third order coefficient also agrees with the value $0.030$ computed in \cite{baker1971singularity}.

Second, around unitarity the EoS expands as
\begin{equation}\label{around_unitarity}
E=\frac{3}{5}NE_F\left(\xi_s-\zeta\frac{1}{k_Fa}+\ldots\right).
\end{equation}
We find the universal parameter of the unitary $T=0$ superfluid, $\xi_s=0.41(1)$ with $2$ $\%$ accuracy. This value is in agreement with recent calculations and measurements \cite{inguscio2006ultracold}. Our thermodynamic measurement $\zeta=0.93(5)$ can be compared with a recent experimental value $\zeta=0.91(4)$ \cite{vale2010universal} as well as the theoretical value $\zeta=0.95$ \cite{lobo2006pair}, both of them obtained through the study of the pair correlation function. This experimental agreement confirms the remarkable link between the macroscopic thermodynamic properties and the microscopic short-range pair correlations, as shown theoretically in \cite{tan2008large}. 

Third, in the BEC limit the EoS of the superfluid is that of a weakly interacting Bose-Einstein condensate of molecules \cite{leeyang1957many,leyronas2007superfluid}:
\be\label{BEC_limit}                   
E =  \frac{N}{2}E_b+  N\frac{\pi\hbar^2a_{dd}}{2m}n\left(1+\frac{128}{15\sqrt{\pi}}\sqrt{n a_{dd}^3}+...\right),                      
\ee
where $a_{dd}=0.6a$ is the dimer-dimer scattering length \cite{inguscio2006ultracold} and $n$ is the dimer density. The term in $\sqrt{n a_{dd}^3}$ is the well-known Lee-Huang-Yang (LHY) correction to the mean-field interaction between molecules \cite{leeyang1957many,leyronas2007superfluid}. Signatures of beyond mean-field effects were previously observed through a pioneering study of collective modes \cite{altmeyer2007precision} and density profile analysis \cite{shin2008rsi} but no quantitative comparison with (\ref{BEC_limit}) was made. Fitting our data in the deep BEC regime with Eq.(\ref{BEC_limit}), we measure the bosonic LHY coefficient 4.4(5), in agreement with the exact value $128/15\sqrt{\pi}\simeq4.81$ calculated for elementary bosons in \cite{leeyang1957many} and recently for composite bosons in \cite{leyronas2007superfluid}. 

Having checked this important beyond mean-field contribution, we can go one step further in the expansion. The analogy with point-like bosons suggests to write the next term as $\left[\frac{8}{3}(4\pi-3\sqrt3)n a_{dd}^3(\log(n a_{dd}^3)+B)\right]$ \cite{sciencesom,wu1959ground,braaten2002dilute}. Using $h_S^{\mathrm{BEC}}(\widetilde{\delta})$ (Fig.\ref{fig_h0}, and \cite{sciencesom}), we deduce the effective three-body parameter for composite bosons $B=7(1)$. Interestingly, this value is close to the bosonic hard-sphere calculation $B=8.5$ \cite{tan2008three} and to the value $B\approx7.2$ for point-like bosons with large scattering length \cite{braaten2002dilute}.
 
\begin{figure}[h!t!]
\centerline{\includegraphics[width=\columnwidth]{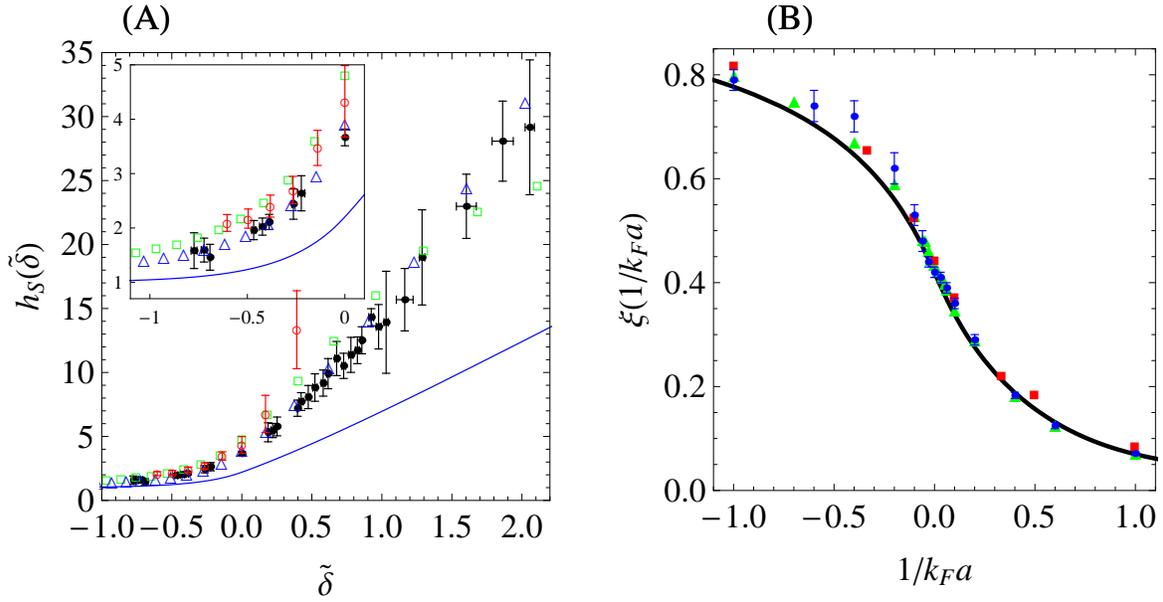}}
\caption{Comparison with many-body theories. (A): Direct comparison of $h_S(\widetilde{\delta})$ with a quantum Monte-Carlo calculation, red open circles \cite{bulgac2008quantum}, a diagrammatic method, green open squares \cite{haussmann2007thermodynamics}, a Nozi\`eres-Schmitt-Rink approximation, blue open triangles \cite{hu2006equation} and the BCS mean-field theory, solid blue line. Inset: Zoom on the BCS side. (B) EoS in the canonical ensemble $\xi(1/k_Fa)$ (solid black line) deduced from the Pad\'e-type approximants to the experimental data $h_S^{\mathrm{BCS}}$ and $h_S^{\mathrm{BEC}}$ plotted in Fig.\ref{fig_h0}. Fixed-Node Monte-Carlo theories: red squares \cite{chang2004quantum}, blue circles \cite{astrakharchik2004eq}, green triangles \cite{pilati2008psi}.}\label{comparison}
\end{figure}

Our measurements also allow direct comparison with advanced many-body theories developed for homogeneous gases in the strongly correlated regime. As displayed in Fig.\ref{comparison}A, our data are in agreement with a Nozi\`eres-Schmitt-Rink approximation \cite{hu2006equation} but shows significant differences from a quantum Monte-Carlo calculation \cite{bulgac2008quantum} and a diagrammatic approach \cite{haussmann2007thermodynamics}. The measured EoS strongly disfavors the prediction of BCS mean-field theory. 

Comparison with Fixed-Node Monte-Carlo theories requires the calculation of the EoS $\xi(1/k_Fa)$ in the canonical ensemble:
\[
\xi\left(\frac{1}{k_Fa}\right)\equiv\frac{E-\frac{N}{2}E_b}{\frac{3}{5}NE_F},
\]
that is deduced from $h_S^{\mathrm{BCS}}(\widetilde{\delta})$ and $h_S^{\mathrm{BEC}}(\widetilde{\delta})$ \cite{sciencesom}. As shown in Fig.\ref{comparison}B, the agreement with theories \cite{chang2004quantum,astrakharchik2004eq,pilati2008psi} is very good. 

We now discuss the EoS of the partially polarized normal phase (black points in Fig.\ref{fig_h}). At low concentrations, we expect the minority atoms to behave as non-interacting quasi-particles, the fermionic polarons \cite{lobo2006nsp}. The polarons are dressed by the majority Fermi sea through a renormalized chemical potential $\mu_2-A(\delta_1)\mu_1$ \cite{schirotzek2009ofp} and an effective mass $m^*(\delta_1)$ \cite{combescot2007nsh,prokof'ev08fpb,pilati2008psi}. Following a Fermi liquid picture, we propose to express the gas pressure as the sum of the Fermi pressure of the bare majority atoms and of the polarons \cite{nascimbene2009eos}:
\begin{equation}
h(\delta_1,\eta)=1+\left(\frac{m^*(\delta_1)}{m}\right)^{3/2}(\eta-A(\delta_1))^{5/2}.\label{ideal_gas}
\end{equation}
Our measured EoS agrees with this model at unitarity and on the BEC side of the resonance (Fig.\ref{fig_h}), where we use for $m^*(\delta_1)$ the most advanced calculations \cite{prokof'ev08fpb,combescot2009analytical}. On the BCS side of the resonance however, we observe at large minority concentrations an intriguing deviation to (\ref{ideal_gas}). In the BCS regime, the superfluid is less robust to spin imbalance. Consequently, the ratio of the two densities $n_1/n_2$ in the normal phase becomes close to unity near the superfluid/normal boundary $\eta_c$. The polaron ideal gas picture then fails. 

\begin{figure}[h!t!]
\centerline{\includegraphics[width=\columnwidth]{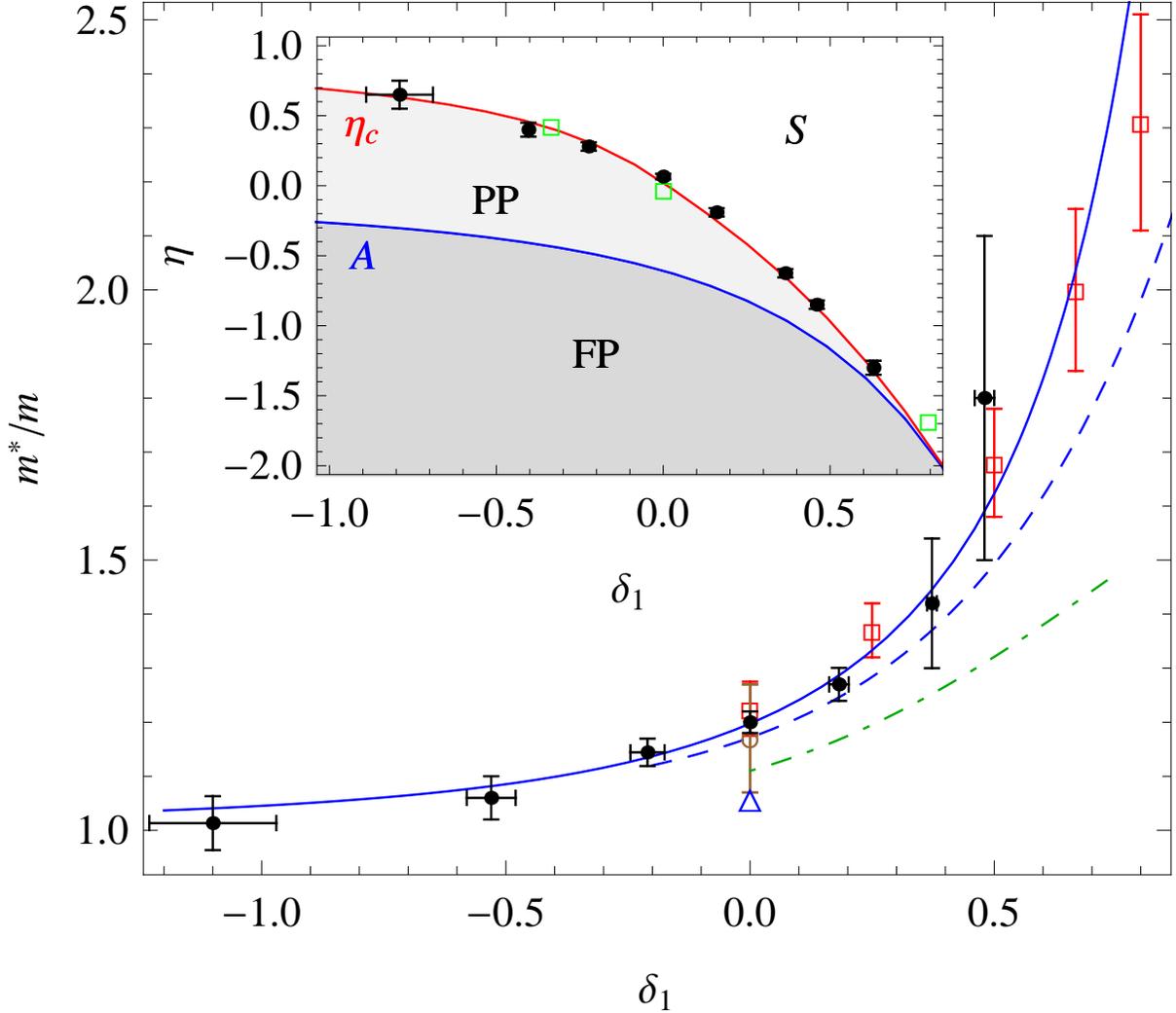}}
\caption{Effective mass $m^*/m$ of the polaron in the BEC-BCS crossover (black dots). The blue dashed line is a calculation from \cite{combescot2007nsh}, red open squares \cite{prokof'ev08fpb}, green dot-dashed line \cite{pilati2008psi}, and blue solid line \cite{combescot2009analytical}. Measurements at unitarity through density profile analysis (blue triangle \cite{shin2008des}) and collective modes study (brown empty circle \cite{nascimbene2009pol}) are also displayed. Inset: Phase diagram of a zero-temperature imbalanced Fermi gas in the BEC-BCS crossover. The blue line is the theoretical value of $A$ \cite{combescot2007nsh,pilati2008psi,prokof'ev08fpb} that sets the separation between the partially polarized (PP) and the fully polarized (FP) phases. Black dots are the measured values of $\eta_c$ (as in Fig.\ref{fig_h}A) which set the separation between the superfluid (S) phase and the partially polarized phase. The red line is the calculation of $\eta_c$ using our EoS of the superfluid and the model (\ref{ideal_gas}) for the normal phase. The green squares are lower bounds of $\eta_c$ given by the values of the gap measured in \cite{schirotzek2008determination}, see \cite{sciencesom}.}\label{eta_delta}
\end{figure}

Alternatively, we can let the effective mass $m^*$ be a free parameter in model (\ref{ideal_gas}) in the fit of our data around $\eta=A$. We obtain the value of the polaron effective mass in the BEC-BCS crossover (Fig.\ref{eta_delta}). 

An important consistency check of our study is provided by the comparison between our direct measurements of $\eta_c(\delta_1)$ (from Fig.\ref{fig_h}, black dots in the inset of Fig.\ref{eta_delta}) and a calculated $\eta_c(\delta_1)$ from Eq.(\ref{ideal_gas}) and the EoS of the superfluid $h_S$. Assuming negligible surface tension, the normal/superfluid boundary is given by equating the pressure and chemical potential in the two phases. This procedure leads to the solid red line in the inset of Fig.\ref{eta_delta}, in excellent agreement with the direct measurements. In addition, by integrating our measured EoS of the homogeneous gas over the trap, one retrieves the critical polarization for superfluidity of a trapped gas, in agreement with most previous measurements \cite{sciencesom}. 

We have measured the equation of state of a two-component Fermi gas at zero temperature in the BEC-BCS crossover. As a first extension, we could explore the thermodynamics of the far BEC region of the phase diagram where a new phase associated with a polarized superfluid appears \cite{shin2008rsi,pilati2008psi}. Another interesting direction would be the mapping of the EoS as a function of temperature and investigate the influence of finite effective range that is playing a key role in higher density parts of neutron stars. 

\newpage
\bibliographystyle{Science}

\noindent{\bf Supporting Online Material}\\
www.sciencemag.org\\
Materials and Methods \\ 
Figures S1, S2, S3, S4\\
References

\newpage

\newpage

\newpage

\newpage

\cleardoublepage

\baselineskip24pt

\bigskip
\noindent{Supporting Online Material for}\\
\noindent{\textbf{\large{The Equation of State of a Low-Temperature Fermi Gas with Tunable Interactions}}\\
\noindent{N. Navon, S. Nascimb\`ene, F. Chevy, C. Salomon}\\

\noindent{\textbf{\large{Materials and Methods}}\\

\noindent\textbf{Extracting the EoS of the uniform gas from the density profiles of trapped clouds}\\
In this section, we provide additional insight on the reconstruction of the EoS of the uniform gas. Several steps are required to deduce the EoS from the doubly-integrated profiles. First, the determination of $(\delta_1,\eta)$ along the $z$ axis requires a precise calibration of $\omega_z$ and the knowledge of the central chemical potentials $\mu_{i}^0$. The axial confinement is produced by a magnetic field curvature, which ensures very good reproducibility. $\omega_z$ is calibrated (to $<1\%$) by measuring the frequency of the axial center of mass dipole mode. $\mu_{1}^0$ is determined using the fully polarized outer rim of the cloud. In this region the density profile is fitted by a Thomas-Fermi formula $\overline{n}_1(z)=\alpha(1-z^2/R_1^2)^{5/2}$, which gives $\mu_1^0=\frac{1}{2}m\omega_z^2R_1^2$. $h$ is then directly obtained by taking the ratio $(\overline{n}_1(z)+\overline{n}_2(z))/\alpha(1-z^2/R_1^2)^{5/2}$, thus avoiding the measurement of the radial frequency $\omega_r$ and cancelling many systematic effects such as imperfect atom counting \cite{shin2008des,nascimbene2009eos}.

The determination of $\mu_2^0$ requires some information on the EoS. We use the outer radius of the minority component as a reference. Indeed, in the limit of vanishing minority spin density, the minority chemical potential is equal to the one of a single minority atom immersed in a Fermi sea of majority atoms, the so-called polaron problem. All advanced calculations \cite{combescot2007nsh,pilati2008psi,prokof'ev08fpb} and the measurement in \cite{schirotzek2009ofp} agree on the value of the chemical potential ratio $A(\delta_1)=\mu_2/\mu_1$ when $n_2\rightarrow0$, which is plotted in the inset of Fig.4. $\mu_2^0$ is then fitted on each image so that $\eta=\mu_{2z}/\mu_{1z}$ is equal to $A(\delta_1(\mu_{1z}))$ at the point $z$ where the minority density vanishes.

Within the local density approximation (LDA), the local chemical potentials along the $z$-axis vary as $\mu_{iz}=\mu^0_i-\frac{1}{2}m\omega_z^2z^2$ (for species $i$). The local interaction parameter, defined as $\delta_{1z}=\hbar/\sqrt{2m\mu_{1z}}a$ (where $1$ is the majority component), also varies along the cloud, as well as the local chemical potential imbalance $\eta_z=\mu_{2z}/\mu_{1z}$. Substituting $z$ in $\delta_1$ in favor of $\eta$, we find:
\be\label{localeta}
\delta_1(\eta)=\delta^0_1\sqrt{\frac{1-\eta}{1-\eta_0}},
\ee
where $\delta^0_1=\hbar/\sqrt{2m\mu_1^0}a$ (resp. $\eta_0=\mu_2^0/\mu_1^0$) is the interaction strength (resp. local imbalance) at the center of the cloud and we dropped the $z$ subscript for clarity. Each trapped density profile thus gives the EoS along the parametric curve $(\delta_1(\eta),\eta)$.

The images used to reconstruct the EoS at a given magnetic field do not perfectly belong to the same curve $(\delta_1(\eta),\eta)$. Here we quantify the systematic error produced by overlapping (and then averaging) the various images. To do so, we have simulated density profiles using our measured EoS with the distribution of initial parameters $\{(\delta_{1j}^0,\eta^0_j)\}$ (where $j$ is the image index) of the images used and have reproduced the reconstruction process. We then compare this result to the expected EoS corresponding to the mean values of $\{(\delta_{1j}^0,\eta^0_j)\}$. The difference is less than $3$ $\%$. Moreover, the determination of $\mu_2^0$ using $A(\delta_1)$ as a reference (as explained in the text), leads to an additional systematic error. We estimate it to be $4$ $\%$ on the $h_S$ axis of Fig.2. The uncertainty on the imaging system magnification leads to a $5$ $\%$ systematic error on the $\widetilde{\delta}$ axis of Fig.2, while trap anharmonicity effects are expected to be $2\%$.\\

The critical chemical potential $\eta_c$ is extracted from each equation of state $h(\delta_1,\eta)$ obtained at a given magnetic field. We fit the data in a region $\eta\in[\overline{\eta}-0.2,\overline{\eta}+0.2]$ with a continuous function made of two straight segments, and observe that the location of the breaking point is insensitive to the $\overline{\eta}$ value defining the set of points chosen for the fit. This shows that our data supports an abrupt change of slope, and $\eta_c$ is identified as the breaking point.

\smallskip
\noindent\textbf{Parametrization of the Superfluid Equation of State}\\
In order to extract physical parameters in the BEC-BCS crossover from our equation of state and to calculate the canonical EoS, we use a simple parametrization of our data that possesses the correct asymptotic behaviors (at $\widetilde{\delta}\rightarrow\pm\infty$ and $\widetilde{\delta}\rightarrow 0$).

First, on the BCS side $\widetilde{\delta}<0$, we use a Pad\'e-type approximant:
\begin{equation}\label{pade_BCS}
h_S^{\mathrm{BCS}}(\widetilde{\delta})=\frac{\widetilde{\delta}^2+\alpha_1\widetilde{\delta}+\alpha_2}{\widetilde{\delta}^2+\alpha_3\widetilde{\delta}+\alpha_4}.
\end{equation}
Using the mean-field asymptotic behavior $h_S(\widetilde{\delta})\simeq1-5/(3\pi\widetilde{\delta})$ in the BCS regime as a constraint on the $\alpha_i$, a fit of our data for $\widetilde{\delta}<0.2$ with (\ref{pade_BCS}) leads to the $\alpha_i$ coefficients gathered in Table S3.

Second, on the BEC side $\widetilde{\delta}>0$, we capture the behavior in the BEC limit (Eq.(5) in the paper with the additional $\log$ term) using the following formula: 
\begin{equation}\label{pade_BEC}
h_S^{\mathrm{BEC}}(\widetilde{\delta})=\frac{\beta_1+\beta_2\widetilde{\delta}+\beta_3\widetilde{\delta}\log(1+\widetilde{\delta})+\beta_4\widetilde{\delta}^2+\beta_5\widetilde{\delta}^3}{1+\beta_6\widetilde{\delta}^2}.
\end{equation}
The $\beta_i$ being contrained by the values of $\xi_s$ and $\zeta$ previously determined from the BCS side and by the exactly known coefficients in (5), we fit our data for $\widetilde{\delta}>-0.2$ with a single free parameter in (\ref{pade_BEC}) and obtain the values of $\beta_i$ in Table S3. \\

\smallskip
\noindent\textbf{The three-body $B$ parameter}\\
For a weakly interacting Bose-Einstein condensate, the ground state energy can be written as \cite{wu1959ground,braaten2002dilute}:          
\be\label{BEC_limit_S}         
E=N\frac{2\pi\hbar^2a}{m}n\left(1+\frac{128}{15\sqrt{\pi}}\sqrt{n a^3}+\left[\frac{8}{3}(4\pi-3\sqrt3)n a^3(\log(n a^3)+B)\right]+...\right),                     
\ee
The first term is the mean-field contribution, the second is the Lee-Huang-Yang correction. The third term involves the three-body problem and was first calculated in \cite{wu1959ground}. It was shown in \cite{leyronas2007superfluid} that this equation of state is rigorously valid for composite bosons up to the LHY term by substracting to the energy the binding energy of the molecules, replacing $a$ by the dimer-dimer scattering length $a_{dd}$, $m$ by the dimer mass and considering $n$ as the dimer density. The analogy with point-like bosons suggests to write the next term for a Bose-Einstein condensate of dimers as $\left[\frac{8}{3}(4\pi-3\sqrt3)n a_{dd}^3(\log(n a_{dd}^3)+B)\right]$. The coefficient before the $\log$ term is given by the low-momenta physics \cite{wu1959ground} where the composite nature of the dimers should not play a role. The coefficient $B$ also involves the three-boson problem at high momenta, unveiling the inner structure of the dimers. For elementary bosons, $B$ depends on the microscopic details of the interaction potential between bosons and involves Efimov physics \cite{braaten2002dilute}. On the other hand, the internal structure of the dimers is completely characterized by the scattering length $a$. Hence, we could expect the three-dimer problem to be solely described by $a$ as well. $B$ would take a universal value, \emph{i.e.} independent on the fermionic species. Using $h_S^{\mathrm{BEC}}(\widetilde{\delta})$, we deduce the effective three-body parameter for composite bosons $B=7(1)$. Interestingly, this value is close to the bosonic hard-sphere calculation $B=8.5$ \cite{tan2008three} and to the value $B\approx7.2$ for point-like bosons with large scattering length \cite{braaten2002dilute}. \\

\smallskip
\noindent\textbf{Derivation of the pressure formula (2)}\\
We consider a mixture of species $i$, of mass
$m_i$, trapped in a harmonic trap of transverse frequencies
$\omega_{ri}$. Using Gibbs-Duhem relation at a constant temperature $T$,
$\textrm{d}P=\sum_in_i \textrm{d}\mu_i$, then
$$
\sum_i\frac{m_i\omega_{ri}^2}{2\pi}\overline{n}_i=\int\sum_i
\frac{m_i\omega_{ri}^2}{2\pi}\textrm{d}x\textrm{d}y \frac{\partial
P}{\partial\mu_i}=\int \sum_i \textrm{d}\mu_i\frac{\partial
P}{\partial \mu_i},
$$
where we have used local density approximation ($\mu_i(\mathbf{r})=\mu_i^0-V(\mathbf{r})$) to
convert the integral over space to an integral on the chemical
potentials. The integral is straightforward and yields
\begin{equation}\label{pngen}
P(\mu_{iz},T)=\frac{1}{2\pi}\sum_im_i\omega_{ri}^2\overline{n}_i(z).
\end{equation}
Eq (2) of the main text is a special case of Eq (\ref{pngen}) for two spin components of equal masses confined by the same trapping potential.  \\

\smallskip
\noindent\textbf{Critical Polarization for a Trapped Gas}\\
Here we compare our measurements of the chemical potential imbalance for the normal/superfluid transition to previous works \cite{shin2006observation,zwierlein2006fsi,nascimbene2009pol,partridge2006deformation}, where this phase transition was characterized by measuring the maximum polarization $P_c$ above which the superfluid is no longer present. We define the polarization $P=(N_1-N_2)/(N_1+N_2)$, where $N_i$ is the total atom number for the species $i$. 
Modeling the normal phase using 
\begin{equation}
h(\delta_1,\eta)=1+\left(\frac{m^*(\delta_1)}{m}\right)^{3/2}(\eta-A(\delta_1))^{5/2},\label{ideal_gas_2}
\end{equation}
we calculate the atom number $N_i$ by integrating the densities $n_i=\partial P/\partial\mu_i$ over the trap: 
\[
N_i=\int\mathrm{d}^3\mathbf{r}\,n_i(\mu_1^0-V(\mathbf{r}),\mu_2^0-V(\mathbf{r}),a).
\]
The critical polarization is obtained when the transition normal to superfluid occurs at the bottom of the trap, \emph{i.e.} for $\mu_2^0/\mu_1^0=\eta_c(\delta_1^0)$, where $\delta_1^0=\hbar/\sqrt{2m\mu_1^0}a$. $\eta_c(\delta_1)$ is provided by the solid red line in Fig.3 (see text). In Fig.S1, we plot $P_c$ as a function of $1/k_Fa$, where $k_F=\hbar(\omega_z\omega_r^2)^{1/3}(6N_1)^{1/3}$, providing a direct comparison with previous experimental data \cite{shin2006observation,zwierlein2006fsi,nascimbene2009pol,partridge2006deformation}. Our results are in excellent agreement with \cite{shin2006observation,zwierlein2006fsi,nascimbene2009pol} but not with \cite{partridge2006deformation}, where the partially polarized phase is absent. As the atom numbers and trap anisotropy in our experiment are close to those in \cite{partridge2006deformation}, this discrepancy remains to be understood. \\

\smallskip
\noindent\textbf{Grand Canonical-Canonical Correspondence}\\
In this section, we explicit the conversion formulas between the grand-canonical and canonical ensembles. We recall that the energy density $\mathcal{E}=E/V$ is linked to the pressure by the usual Legendre transform $\mathcal{E}=-P+\mu n$ where $\mu=\frac{\mu_1+\mu_2}{2}$ is the chemical potential in the fully paired superfluid and $n$ is the total atomic density. Combining this relationship with the Gibbs-Duhem relation $dP=nd\mu$ and using the proper normalization for the dimensionless function $h(\delta)$, one finds the two formulas:
\begin{eqnarray}
x(\delta)&=&\frac{\delta}{(h(\delta)-\frac{\delta}{5}h'(\delta))^{1/3}}\label{xdelta}\\
\xi(\delta)&=&\frac{h(\delta)-\frac{\delta}{3}h'(\delta)}{(h(\delta)-\frac{\delta}{5}h'(\delta))^{5/3}}\label{xidelta}
\end{eqnarray}
Eq (\ref{xdelta}) relates the natural variables of both ensembles, while Eq (\ref{xidelta}) provides the canonical EoS. The canonical EoS (displayed in Fig.3b) is then simply obtained through a parametric plot of $(x(\delta),\xi(\delta))$ using the Pad\'e-type fitting functions for $h(\delta)$. Another familiar quantity that can be calculated from our fits is the chemical potential in the BEC-BCS crossover:
\be
\frac{\mu}{E_F}=x(\delta)^2\left(\frac{1}{\delta^2}-\theta(\delta)\right),
\ee 
where $\theta(\delta)$ is the Heaviside step function.\\

\smallskip
\noindent\textbf{Thermometry}\\
We perform thermometry of our imbalanced gases on the fully polarized outer shell of the cloud, which is non-interacting. We thus fit the wings of the density profiles with finite-temperature Thomas-Fermi distributions \cite{shin2008phase}. Using this technique, we find a temperature of $T/T_F=0.03(3)$ at the unitary limit and on the BEC side of the resonance, a temperature much smaller than the critical temperature for superfluidity $T_c$ in this interaction range, justifying the $T=0$ assumption.

On the BCS side, the small size of the fully polarized region renders this method inaccurate and gives an experimental upper bound $T/T_F<0.13$. However, the observation of a wide superfluid plateau on the doubly-integrated density difference is a signature of an unpolarized superfluid phase clearly indicating that $T<T_c$. In the normal phase, finite temperature corrections are expected to be of the order of $(T/T_F)^2$. As $T<T_c\ll T_F$ in the BCS regime, they are expected to be negligible. In addition, we observe an abrupt change of slope at the normal/superfluid boundary indicating the presence of a first-order phase transition. Consequently, this gives an upper bound $T<T_\mathrm{tri}$ where $T_\mathrm{tri}$ is the temperature of the tri-critical point at which the phase transition becomes of second order. Finite-temperature corrections are expected to be dominated by thermal excitations of Bogoliubov-Anderson phonons. From the superfluid equation of state determined from our data we compute the speed of sound $c=\sqrt{n/m\,\partial\mu/\partial n}$ as a function of interaction strength, and the finite-temperature correction to the pressure for $T=T_\mathrm{tri}$ predicted by a mean-field theory \cite{buzdin1997generalized,combescot2002transition,parish2007finite}. We thus infer that the systematic error on $h_S$ due to a finite temperature is less than 2~$\%$ for our data on the BCS side of the resonance.\\

\smallskip
\noindent\textbf{Link between the critical imbalance $\eta_c$ and the pairing gap $\Delta$}\\
An intriguing link exists between the measurements of $\eta_c(\delta_1)$ and the superfluid pairing gap $\Delta$. Indeed, for $\delta_1<0.6$ we have observed that the superfluid is unpolarized. An unpolarized superfluid becomes unstable as soon as flipping the spin of a minority atom decreases the grand-potential $\Omega=E-\mu_1N_1-\mu_2N_2$. After the spin-flip, a pair is broken and releases an energy equals to $\Delta E=2\Delta$. Thus the total grand-potential difference is $\Delta\Omega=\mu_2-\mu_1+2\Delta$. The superfluid is stable against infinitesimal polarization as long as $\Delta\Omega>0$, hence the minority chemical potential $\mu_2$ must be lower than $\mu-\Delta$, where $\mu=(\mu_1+\mu_2)/2$. This argument yields lower bounds on the values of $\eta_c$: $\eta_c>(1-\Delta/\mu)/(1+\Delta/\mu)$ (empty green squares in the inset of Fig.4). We observe that our $\eta_c$ data is very close to these bounds, calculated from previous experimental determinations of the gap from \cite{schirotzek2008determination}. As initially pointed out in \cite{carlson2005atc} for the unitary gas, this suggests that in the investigated range the superfluid to normal transition is driven by single-particle excitations of the superfluid.

\newpage

\newpage
\begin{figure}[h!]
\centerline{\includegraphics[width=\columnwidth]{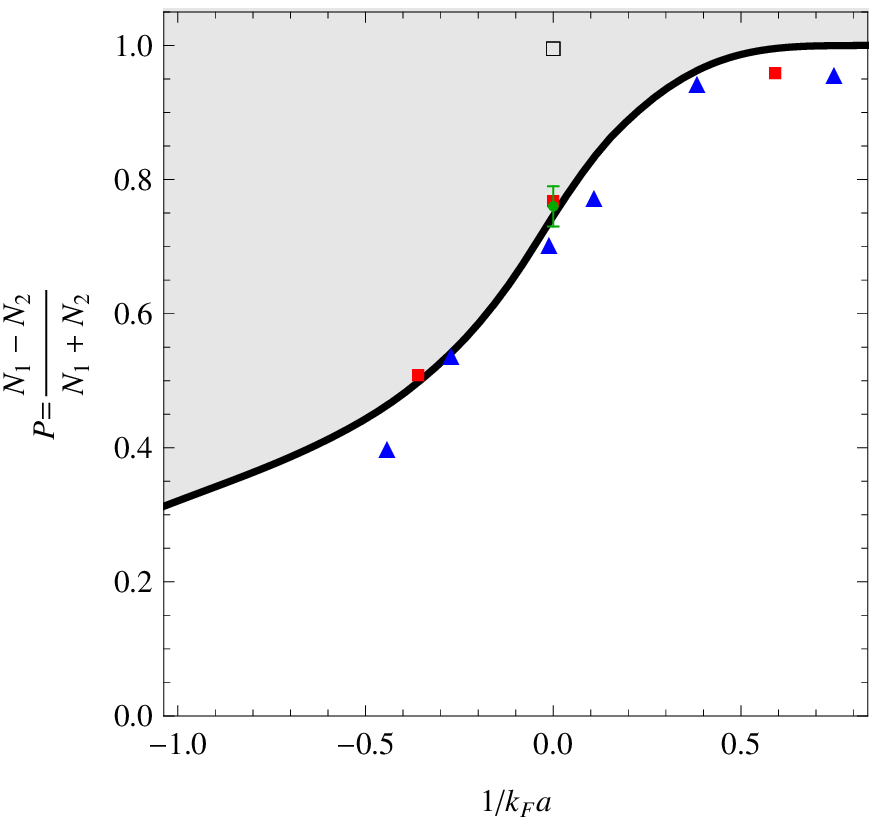}}
\caption*{\textbf{Fig. S1.} Critical Polarization $P_c$ of a trapped imbalanced Fermi gas in the BEC-BCS crossover (solid black line) and comparison with the data from MIT (red squares \cite{shin2006observation} and blue triangles \cite{zwierlein2006fsi}), ENS (green circle \cite{nascimbene2009pol}), and Rice (empty black square \cite{partridge2006deformation}). In the gray region $P>P_c$, the superfluid phase is absent.}\label{Pc}
\end{figure}

\newpage
\begin{table}[h!]
  \centering
  \begin{tabular}{cccccc}
              \hline
              \hline
              &$\alpha_1$   & $\alpha_2$     & $\alpha_3$ & $\alpha_4$ &    \\
              &-1.137& 0.533  &    -0.606     & 0.141       &     \\
              \hline
              $\beta_1$ & $\beta_2$& $\beta_3$ & $\beta_4$ & $\beta_5$ &$\beta_6$ \\
              3.78 & 8.22& 8.22 & -4.21&3.65& 0.186\\
              \hline
              \hline
  \end{tabular}
  \caption*{\textbf{Table S2.} Pad\'e-type approximants coefficients $\alpha_i$ and $\beta_i$ fitted from our data.}\label{coeffs}
\end{table} 

\newpage
\begin{figure}[h!]
\centerline{\includegraphics[width=\columnwidth]{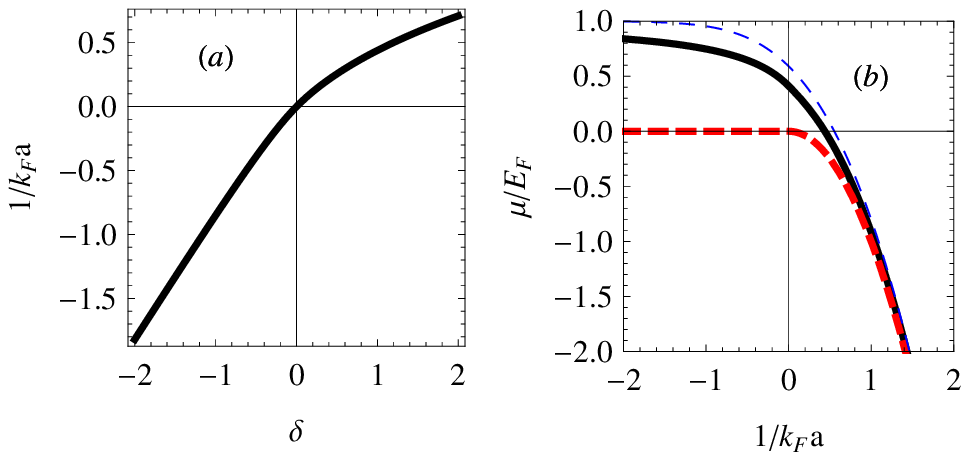}}
\caption*{\textbf{Fig. S3.} Canonical-Grand Canonical Correspondence in the BEC-BCS crossover. (a) Canonical natural variable $1/k_Fa$ as a function of the grand-canonical natural variable $\delta$. (b) Chemical potential in the canonical ensemble (black solid line). For comparison, mean-field BCS theory is the dashed blue line, the binding energy, the dashed red line.}\label{gcc}
\end{figure}

\newpage
\begin{figure}[h!]
\centerline{\includegraphics[width=\columnwidth]{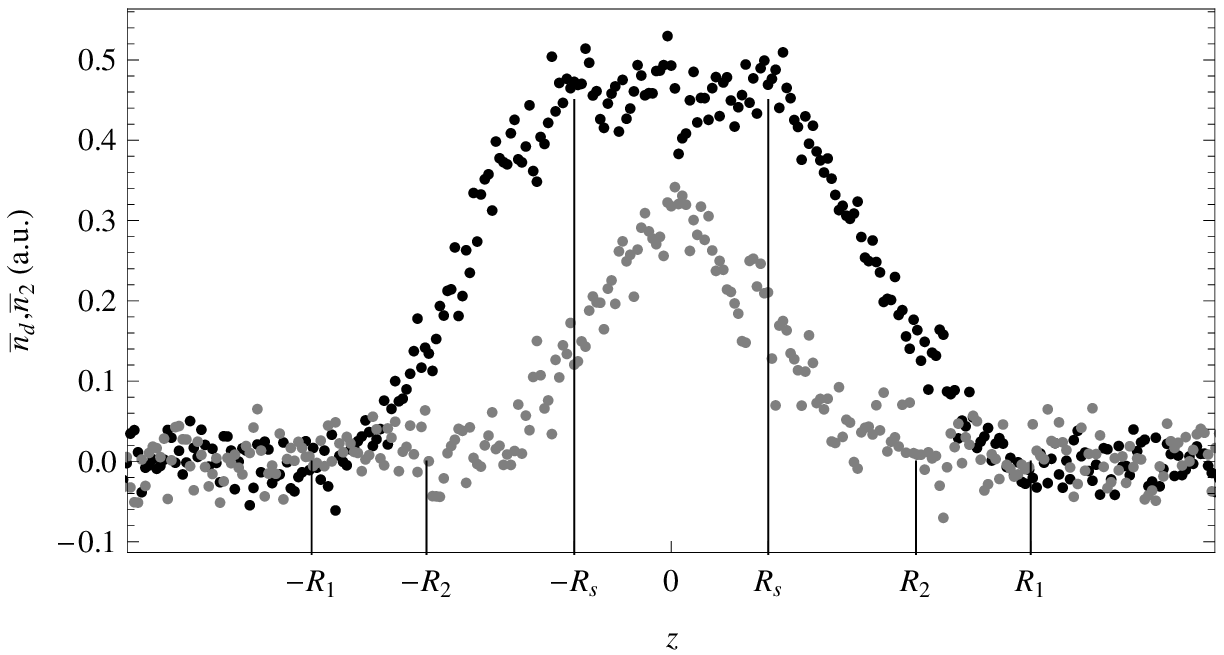}}
\caption*{\textbf{Fig. S4.} Raw data: doubly-integrated density profiles $\overline{n}_2(z)$ (gray dots) and $\overline{n}_d(z)=\overline{n}_1(z)-\overline{n}_2(z)$ (black dots) for a gas prepared in the unitary limit. $R_1$, $R_2$, $R_S$ are the boundaries of the fully polarized phase, of the partially polarized phase, and of the superfluid core, respectively. The plateau on the density difference $\overline{n}_d(z)$ observed in the region $|z|<R_S$ indicates equal densities for both spin components in the superfluid phase.}\label{FigS4}
\end{figure}


\newpage
\bibliographystyle{Science}

\end{document}